\documentclass[conference]{IEEEtran}
\usepackage{ragged2e}
\usepackage{floatrow}
\usepackage{amsthm}
\usepackage[utf8]{inputenc}
\usepackage[english]{babel}
\theoremstyle{remark}

\usepackage{multicol}
\usepackage{epsfig}
\usepackage{epstopdf}
\usepackage{float}
\usepackage{perpage}
\MakeSorted{figure}
\MakeSorted{table}
\usepackage{array}
\usepackage{graphicx}
\usepackage{amsfonts}
\usepackage{amssymb}

\usepackage{enumitem}
\usepackage[english]{babel}
\usepackage[utf8]{inputenc}
\usepackage[linesnumbered,ruled,vlined]{algorithm2e}

\usepackage{cite}
\DeclareMathAlphabet\mathbfcal{OMS}{cmsy}{b}{n}
\ifCLASSINFOpdf
\else
\fi
\usepackage[cmex10]{amsmath}
\usepackage{array}

\usepackage[linesnumbered,ruled,vlined]{algorithm2e}
\usepackage[font=footnotesize,labelfont=bf, figurename=Fig.]{caption} 
\usepackage{subcaption}
\usepackage{fix-cm}
\usepackage{etoolbox} 
\newtoggle{SINGLE_COL}
\toggletrue{SINGLE_COL}     
\togglefalse{SINGLE_COL}   

\newtoggle{BIG_EQUATION}
\toggletrue{BIG_EQUATION}  

\IEEEoverridecommandlockouts
\IEEEpubid{\makebox[\columnwidth]{978-1-5386-3821-7/18/\$31.00~\copyright~2018~IEEE \hfill}
\hspace{\columnsep}\makebox[\columnwidth]{ }}

 \author{
\IEEEauthorblockN{Arpita Singh Chauhan, Ekant Sharma, and Rohit Budhiraja }
\IEEEauthorblockA{Dept. of Electrical Engineering, 
 IIT Kanpur, Kanpur, 208016\\
 Email: arpitach@iitk.ac.in, ekant@iitk.ac.in, and rohitbr@iitk.ac.in}
 }

\begin{document}
\title{Hybrid Block Diagonalization for Massive MIMO Two-Way Half-Duplex AF Hybrid Relay}
\maketitle

\maketitle
\begin{abstract}
We consider a multi-pair two-way amplify-and-forward massive multi-input multi-output (MIMO) hybrid relay with MIMO user-pairs. A hybrid relay has lesser number of radio frequency (RF) chains than the antennas, which significantly reduces the implementation cost. We employ block-diagonalization-based baseband processing at the hybrid relay to cancel the inter user-pair interference and equal-gain-combining-based RF processing to maximize the beamforming gain. We also use an algebraic norm maximizing relay transmit strategy to maximize the spectral efficiency (SE) of each user-pair. We numerically show that the proposed hybrid relay has only marginally inferior SE than a full RF-chain relay.
\end{abstract}
\IEEEpeerreviewmaketitle

\section{Introduction}
Cooperative communication is widely investigated to extend the coverage, reduce the power consumption and improve the spectral efficiency (SE) of wireless networks \cite{Lee_Hanzo}. A one-way half-duplex (HD) relay requires four channel uses for information exchange between a user-pair. A two-way HD relay requires only two channel uses for the information exchange \cite{Lee_Hanzo}, yielding much higher SE than a one-way HD relay. Two-way HD relaying is also extended to enable multiple user-pairs exchange information in two channel uses~\cite{CuiSJ14}. The SE of multi-pair two-way HD relaying can be further boosted by incorporating massive multi-input multi-output (MIMO) technology in a relay~\cite{CuiSJ14}.


In conventional massive MIMO relaying, the number of radio-frequency (RF) chains at the relay is equal to the number of antennas, which leads to a very high implementation cost, computational complexity and energy consumption \cite{LiangXD14}. References \cite{LiangXD14, NiD16,tcomXuLJD17} uses a hybrid transceiver architecture to reduce the number of RF chains by connecting each RF chain to multiple antennas. For such a hybrid architecture, the relay processing is divided into two parts: i) high-dimensional RF processing, using variable analog phase shifters, which modifies only the phase of the complex signals; and ii) low-dimensional digital baseband processing which modifies both amplitude and phase of the complex signals. 

Reference \cite{LiangXD14} designed an RF beamformer for a hybrid single-hop massive MIMO system, based on equal-gain combining (EGC) principles~\cite{LoveH03}, to realize large beamforming gain. Reference \cite{NiD16} considered a hybrid massive MIMO system with MIMO users, and  designed block-diagonalization-based low-dimensional digital precoder to cancel inter user-pair interference, and EGC-based RF beamformer to realize beamforming gain. Reference \cite{tcomXuLJD17} considered a massive MIMO decode-and-forward (DF) hybrid HD relay with single-antenna users, and designed a zero-forcing-based digital precoder and a EGC-based RF beamformer. 

This paper extends the work of \cite{tcomXuLJD17}, and designs digital baseband and analog RF beamformers for an amplify-and-forward (AF) hybrid two-way HD relaying with multiple MIMO user-pairs. We note that the design in \cite{tcomXuLJD17} for a DF relay with single-antenna users is not applicable to the AF relay system with MIMO users considered herein because of the i) coupling of channels in an AF relay; and ii) MIMO users.  To the best of our knowledge, beamforming design for hybrid two-way massive MIMO relaying with MIMO users has not yet been investigated in the literature. We next list the \textbf{main} contributions of this~paper.

1) We construct a block-diagonalization-based digital baseband precoder at the relay to cancel the inter user-pair interference. The proposed design decomposes the multiple user-pairs two-way HD relaying system into multiple non-interfering single user-pair two-way relay systems by cancelling the inter user-pair interference.  We then design relay amplification matrix by employing an algorithm proposed in \cite{RoemerH09} that optimizes the SE of each user-pair by maximizing the algebraic norm of channels between them. We also design an EGC-based RF analog beamformer to realize the beamforming gains of the massive MIMO relay. 

2) We numerically show that a hybrid relay provides similar SE as that of a full-RF chain relay, with much lower implementation complexity.

\textit{\!\!\!\!\!\!Notations}: The boldface upper-case and lower-case letters represent the matrices and vectors, respectively. The notations $\mathbb{E}[\cdot]$, $\mbox{Tr}\lbrace\cdot\rbrace,(\cdot)^{\ast},(\cdot)^{T},(\cdot)^{H},(\cdot)^{-1}$ represents the expectation, trace, complex-conjugate, transpose, Hermitian,  and inverse operations. The operator $\Vert\!\! \cdot \!\!\Vert$ denotes the Euclidean norm of a vector and $\|\! \cdot \!\|_{F}$ represents the  Frobenious norm of a matrix. The operator blkdiag$[\cdot]$ creates a block-diagonal matrix by aligning the input matrices onto its diagonal. The operator vec$\{\cdot\}$ stacks the columns of the matrix into a vector. The operator $ \text{unvec}_{a\times b}\{\cdot\} $  converts a vector into an $ a \times b $ matrix. The symbol $\mathbf{I}_{n}$ represents an $n \times n$ identity matrix. The operator $ \otimes$ denotes the Kronecker product.

\section{System Model}\label{sys_model}
We consider a multi-pair two-way AF massive MIMO hybrid HD relay system, where $K$ HD user-pairs communicate via a single HD relay on the same time-frequency resource. We assume that each user has $N_{U}$ antennas, and the relay has $ N_{R} $ antennas and only $M_{R} \ll N_{R}$ RF chains.
Each user transmits $M_{D} \leq N_{U}$ data streams.  We assume that a user pair $(2m-1,2m)$ for $m=1,\cdots,K$ exchanges information and there is no direct link between the users due to large path loss and shadowing \cite{CuiSJ14}.\footnote{To avoid repetition, we assume $m=1,\cdots, K$ and $k=1,\cdots,2K$ throughout this paper.}

Let the information vector for the $k$th user be $\sqrt{\frac{p_{k}}{N_{U}}}\mathbf{s}_{k}\in\mathbb{C}^{M_{D} \times 1}$, which it multiplies with the baseband precoder $\mathbf{D}_{k} \in \mathbb{C}^{N_{U} \times M_{D}}$ to generate the transmit vector 
\begin{align}\label{xk}
\mathbf{x}_{k}= \sqrt{\frac{p_{k}}{N_{U}}}\mathbf{D}_{k} \mathbf{s}_{k}  \in \mathbb{C}^{N_{U} \times 1},
\end{align}
which satisfies the power constraint $ \mathbb{E}\parallel \mathbf{x}_{k}\parallel^{2}\leq p_{k} $. Two-way AF relaying consists of two phases - the multiple access and the broadcast. In the multiple access phase, all the $2K$ users simultaneously transmit their respective signals to the relay, which receives a sum signal given as
\begin{align}\label{r}
\mathbf{y}_R&=\sum_{k=1}^{2K} \mathbf{H}_{k}\mathbf{x}_{k} +\mathbf{n}_{R}= \mathbf{H}\mathbf{x} +\mathbf{n}_{R}.
 \end{align}
Here $\mathbf{x}=\left[\mathbf{x}_{1}^{T},\, \mathbf{x}_{2}^{T},\cdots,\,\mathbf{x}_{2K}^{T}\right]^{T}\in\mathbb{C}^{2KN_{U}\times 1}$, $\mathbf{H}=\left[\mathbf{H}_{1},\,\mathbf{H}_{2},\cdots,\,\mathbf{H}_{2K}\right]\in\mathbb{C}^{N_{R}\times 2KN_{U}}$, where $\mathbf{H}_{k}\in\mathbb{C}^{N_{R}\times N_{U}}$ denotes the channel between the $k$th user and the relay. We assume that the elements of channel matrix $\mathbf{H}$ are independent and identically distributed (i.i.d.) with pdf $\mathcal{CN}(0,1)$. The vector $\mathbf{n}_{R} \in \mathbb{C}^{N_{R} \times 1}\sim \mathcal{CN}(0,\sigma_{R}^{2}\mathbf{I}_{N_{R}})$ represents the additive white Gaussian noise (AWGN) at the relay. 

In the broadcast phase, the relay generates its transmit signal $\mathbf{x}_R$ by multiplying its received signal $\mathbf{y}_R$  with a beamforming matrix $ \mathbf{W} \in \mathbb{C}^{N_{R} \times N_{R}} $ as 
\begin{align}\label{hatr}
\mathbf{x}_R=\mathbf{W} \mathbf{y}_R=\mathbf{WHx}+\mathbf{Wn}_R.
\end{align}
and then broadcasts it to all users. The relay transmit signal satisfies the following power constraint  $\mbox{Tr}\left\{\mathbb{E}\left[{\mathbf{x}_R}{\mathbf{x}_R}^{H}\right]\right\}\le P_R$. We assume that the relay operates in time division duplex (TDD) mode and, therefore, channels between the users and the relay are reciprocal. The signal received at the $k^{'}$th user~is 
\begin{align}\label{yk}
\mathbf{y}_{k^{'}} =&\mathbf{H}_{k^{'}}^{ T}\mathbf{x}_R+\mathbf{n}_{k^{'}}=\mathbf{H}_{k^{'}}^{ T}\mathbf{WHx}+\mathbf{H}_{k^{'}}^{ T}\mathbf{Wn}_R+\mathbf{n}_{k^{'}}\nonumber\\
= & \underset{\text{desired signal}}{\underbrace{\mathbf{H}_{k^{'}}^{ T}\mathbf{W} \mathbf{H}_{k} \mathbf{x}_{k}}} + \underset{\text{self-interference}}{\underbrace{\mathbf{H}_{k^{'}}^{ T} \mathbf{W} \mathbf{H}_{k^{'}}\mathbf{x}_{k^{'}}}} \nonumber \\
 & +\underset{\text{inter user-pair interference}}{\underbrace {\sum_{ \scriptstyle {i} \neq k,k^{'}} \mathbf{H}_{k^{'}}^{T} \mathbf{W} \mathbf{H}_{i} \mathbf{x}_{i}}}   + \underset{\text{amplified relay noise}}{\underbrace{\mathbf{H}_{k^{'}}^{T}\mathbf{W}\mathbf{n}_{R}}}+\mathbf{n}_{k^{'}},  
\end{align}
where $\mathbf{n}_{k^{'}}\sim \mathcal{CN}(0,\sigma_{k^{'}}^{2}\mathbf{I}_{N_{U}})$ represents the AWGN at the $k^{'}$th user.  After self-interference cancellation, the received signal in \eqref{yk} is re-written as
\begin{align}\label{tildeyk'}
\mathbf{\widetilde{y}}_{k^{'}} &= {{\mathbf{H}_{k^{'}}^{ T}\mathbf{W} \mathbf{H}_{k} \mathbf{x}_{k}}} +  { {\sum_{ i \neq k,k^{'}} \mathbf{H}_{k^{'}}^{T} \mathbf{W} \mathbf{H}_{i} \mathbf{x}_{i}}} +\tilde{\mathbf{n}}_{k^{'}},
\end{align}
 where \begin{align}\label{eq_noise}
 \tilde{\mathbf{n}}_{k^{'}}=\mathbf{H}_{k^{'}}^{T}\mathbf{W}\mathbf{n}_{R}+\mathbf{n}_{k^{'}},
 \end{align}
  is the equivalent noise at the receiver.
An estimate of the transmitted signal ${\mathbf{s}}_{k^{'}}$ is obtained using the baseband decoder $\mathbf{Q}_{k^{'}} \in \mathbb{C}^{M_{D} \times N_{U}}$ as follows\begin{align}\label{hatsk}
\mathbf{\hat{s}}_{k^{'}}=\mathbf{Q}_{k^{'}} \mathbf{\widetilde{y}}_{k^{'}}.
\end{align}
The SE for the $k^{'}$th user is \cite{NiD16}

\begin{align}\label{gammak}
\gamma_{k^{'}}\!&=\!\frac{1}{2}\log\!\left(\left\vert\mathbf{I}_{M_{D}}\!+\! \frac{p_{k^{'}}}{N_{U}}\mathbf{R}_{k^{'}}^{-1}\mathbf{Q}_{k^{'}}\mathbf{H}_{k^{'}}^{T}\mathbf{W}\mathbf{H}_{k}\mathbf{D}_{k} \mathbf{D}_{k}^{H}\right.\right.\nonumber\\
&\biggl.\biggl.\hspace{1in}\times\mathbf{H}_{k}^{H}\mathbf{W}^{H}\mathbf{H}_{k^{'}}^{\ast}\mathbf{Q}_{k^{'}}^{H}
\biggr \vert \biggr),
\end{align}
where $\displaystyle\mathbf{R}_{k^{'}}\!=\!\!\!\sum_{i \neq k,k^{'}}\!\!\frac{p_{i}}{N_{U}} \mathbf{Q}_{k^{'}}\mathbf{H}_{k^{'}}^{T}\mathbf{W}\mathbf{H}_{i} \mathbf{D}_{i} \mathbf{D}_{i}^{H}\mathbf{H}_{i}^{H}\mathbf{W}^{H}\mathbf{H}_{k^{'}}^{\ast} \mathbf{Q}_{k^{'}}^{H} + \sigma_{R}^{2} \mathbf{Q}_{k^{'}} \mathbf{H}_{k^{'}}^{T}\mathbf{W}\mathbf{W}^{H} \mathbf{H}_{k^{'}}^{\ast} \mathbf{Q}_{k^{'}}^{H} + \sigma_{k^{'}}^{2}\mathbf{Q}_{k^{'}}\mathbf{Q}_{k^{'}}^{H}$.

The SE (in bps/Hz) of the system is therefore
 \begin{align}\label{SE}
 \mbox{SE}= \sum_{k=1}^{2K}\gamma_{k}.
 \end{align}
\iftoggle{SINGLE_COL}{\begin{figure*}[htbp]}{\begin{figure*}[htbp]}
    \centering
    \includegraphics[width=0.75\linewidth,height=0.2\linewidth]{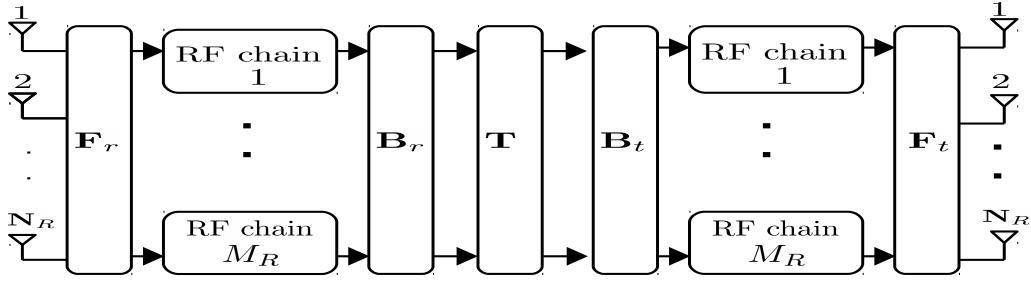}
  \caption{\small Architecture of the relay precoder.}
  \label{system_model}
  \end{figure*}

\section{Proposed design}\label{proposed}


For a massive MIMO relay, consisting of large number of antennas, it is costly to connect each antenna to a separate RF chain{\cite{tcomXuLJD17}}. A hybrid relay, where each RF chain is connected to multiple antennas, can significantly reduce the system cost {\cite{tcomXuLJD17}}. A hybrid massive MIMO transmitter processes its signal in two steps: i) digital amplitude and phase beamforming in digital baseband; and ii) analog phase-only beamforming using RF phase shifters. A hybrid massive MIMO receiver processes its signal in the opposite order. In this work, we consider a massive MIMO hybrid relay for which, as shown in Fig.~\ref{system_model}, we decompose the relay precoder as 
\begin{align}\label{W}
\mathbf{W}=\alpha\widetilde{\mathbf{W}}=\alpha \mathbf{F}_{t} \mathbf{B}_{t} \mathbf{T} \mathbf{B}_{r} \mathbf{F}_{r}.
\end{align} 
Here $ \alpha $ is the amplification factor required to satisfy the relay power constraint. The matrices $\mathbf{F}_{t}\in\mathbb{C}^{N_{R}\times M_{R}}$ and $\mathbf{F}_{r}\in\mathbb{C}^{M_{R}\times N_{R}}$ denote the transmit and receive analog phase-only beamformers, respectively, and  the matrices $\mathbf{B}_{t}\in\mathbb{C}^{M_{R}\times M_{R}}$, $\mathbf{B}_{r}\in\mathbb{C}^{M_{R}\times M_{R}}$ denote the transmit and receive digital precoders, respectively. The matrix $\mathbf{T}\in\mathbb{C}^{M_{R}\times M_{R}}$ denotes the digital relay amplification matrix. We next design these beamformers.


\subsection{Design of analog phase-only beamformers $ \mathbf{F}_{r}$ and $\mathbf{F}_{t}$}
The analog beamformer $ \mathbf{F}_{r} $  is designed to maximize the beamforming gain by employing equal gain combining \cite{LoveH03}:
\begin{align}\label{Fr}
[\mathbf{F}_{r}]_{i,j}=\frac{1}{\sqrt{N_{R}}}e^{j\psi_{i,j}}.
\end{align}
Here $ \psi_{i,j} $ is the phase of $(i,j)$th element of $\mathbf{H}^{H}  $. 
Due to TDD channel reciprocity, we have $\mathbf{F}_{t}=\mathbf{F}_{r}^{T}\in\mathbb{C}^{N_{R}\times 2KN_{U}}$.  We assume, similar to  \cite{NiD16}, that the number of RF chains at the relay is equal to the total number of antennas at all the users i.e., $M_{R}=2KN_{U}$.

\subsection{Design of digital precoders $\mathbf{B}_{r}$ and $\mathbf{B}_{t}$}
We design $\mathbf{B}_{r}$ and $\mathbf{B}_{t}$, using block diagonalization approach {\cite{SpencerSH04}},  to cancel the inter user-pair interference. To this end, we first define an equivalent multiple access-phase channel with analog beamformer as follows
\begin{align}\label{HE}
\mathbf{H}_{E}=  \mathbf{F}_{r}  \left[ \mathbf{H}_{1}, \mathbf{H}_{2},\cdots , \mathbf{H}_{2K}\right]= \left[ \mathbf{\widetilde{H}}_{1}, \mathbf{\widetilde{H}}_{2},\cdots , \mathbf{\widetilde{H}}_{2K}\right]\!,\!
\end{align}
where $\widetilde{\mathbf{H}}_{k}=\mathbf{F}_{r}{\mathbf{H}}_{k}\in\mathbb{C}^{M_{R}\times N_{U}}$ represents the composite channel of the $k$th user.
To design $\mathbf{B}_{r}$, we first define its structure as $\mathbf{B}_{r}=\left[\mathbf{B}_{r1}^{T},\, \mathbf{B}_{r2}^{T},\,\cdots,\,\mathbf{B}_{rK}^{T}\right]^{T}$, where $\mathbf{B}_{rm} \in \mathbb{C}^{2N_{U} \times 2KN_{U}}$ for $m=1,\cdots,K$ is the receive digital precoder for  the $m$th user-pair, and is designed to cancel the inter user-pair interference alone.  
To achieve this objective, we define  $\mathbf{\overline{H}}_{m}\in\mathbb{C}^{2KN_{U} \times 2(K-1)N_{U}}$ as
\begin{align}\label{barHm}
\mathbf{\overline{H}}_{m}\triangleq \left[\! \widetilde{\mathbf{H}}_{1},\cdots\!, \widetilde{\mathbf{H}}_{2m-2},\widetilde{\mathbf{H}}_{2m+1},\cdots\!, \widetilde{\mathbf{H}}_{2K}\!\right],
\end{align}
where $[ \widetilde{\mathbf{H}}_{2m-1},\,\widetilde{\mathbf{H}}_{2m} ] \in \mathbb{C}^{2KN_{U} \times 2N_{U}}$ is the concatenated channel matrix for the $m$th user-pair. To cancel the inter user-pair interference {for the $m$th user-pair}, the digital precoder $\mathbf{B}_{rm}$ should belong to the left null space of $\mathbf{\overline{H}}_{m}$ {\cite{SpencerSH04}}, which can be found  by performing  its singular value decomposition (SVD) as  
\begin{align}\label{svdbarHm}
\mathbf{\overline{H}}_{m} =\left[{\mathbf{\overline{U}}}_{m}, \, \widetilde{\mathbf{U}}_{m}\right] \mathbf{\overline{\Sigma}}_{m}\mathbf{\overline{V}}_{m}^{^{H}}.
\end{align}
Here $ \mathbf{\overline{U}}_{m} $ contains the first $ 2(K-1)N_{U} $ left singular vectors of $ \mathbf{\overline{H}}_{m} $, and $ \widetilde{\mathbf{U}}_{m} $ contains the rest of $ 2N_{U} $ left singular vectors,  which form an orthogonal basis for the left null space of  $\mathbf{\overline{H}}_{m}$. The matrices $ \mathbf{\overline{\Sigma}}_{m} $ and $ \mathbf{\overline{V}}_{m} $ contain the singular values and the right singular vectors of  $\mathbf{\overline{H}}_{m} $, respectively. We now have
\begin{align}\label{tildeUmHj}
\widetilde{\mathbf{U}}_{m}^{H}\mathbf{\widetilde{H}}_{j}= \left\{ \begin{array}{rclc}
\widetilde{\mathbf{U}}_{m}^{H}\mathbf{\widetilde{H}}_{j}, & j=2m-1, 2m\\ 0, & j\neq 2m-1,2m. 
\end{array}\right.
\end{align}
Based on \eqref{tildeUmHj}, we can design the receive digital precoder for $m$th {user-pair} as $\mathbf{B}_{rm}=\widetilde{\mathbf{U}}_{m}^{H} $. 
The equivalent multiple access-phase channel with both analog beamformer and digital precoder is given as 
\begin{align}\label{Hbd}
\mathbf{H}_{BE}&= \mathbf{B}_{r}\mathbf{H}_{E}=\left[\widetilde{\mathbf{U}}_{1}^{*}, \cdots,\widetilde{\mathbf{U}}_{K}^{*} \right]^{T} \left[ \mathbf{\widetilde{H}}_{1}, \cdots , \mathbf{\widetilde{H}}_{2K}\right] \nonumber\\
&=\left[\begin{array}{ccc}\!\mathbf{\widetilde{U}}_{1}^{H}[\mathbf{\widetilde{H}}_{1}, \mathbf{\widetilde{H}}_{2}]  & \cdots & \mathbf{0}  \\ \vdots & \ddots & \vdots \\  \mathbf{0} & \cdots & \mathbf{\widetilde{U}}_{K}^{H}[\mathbf{\widetilde{H}}_{2K-1}, \mathbf{\widetilde{H}}_{2K}]\end{array}\right].
\end{align}
The transmit digital precoder, due to channel reciprocity, is given as $\mathbf{B}_{t}=\mathbf{B}_{r}^{T}$.

The structure of the  matrix $\mathbf{T}=\text{blkdiag}\left[ \mathbf{T}_{1},\cdots,\mathbf{T}_{K}\right]$, where $  \mathbf{T}_{m} \in \mathbb{C}^{2N_{U} \times 2N_{U}}$ is the  relay amplification matrix for the $m$th user pair. Due to the block-diagonal structure of $\mathbf{T}$, the equivalent channel  between the user-pairs can be expressed~as
\begin{align}\label{hatH}\mathbf{\widehat{H}}&= \mathbf{H}_{E}^{T}\mathbf{B}_{t}\mathbf{T}\mathbf{B}_{r}\mathbf{H}_{E} 
=\mbox{blkdiag}\left[\widehat{\mathbf{H}}_{1},\,\widehat{\mathbf{H}}_{2},\cdots,\,\widehat{\mathbf{H}}_{K}\right],
\end{align}
where
\begin{align}
\widehat{\mathbf{H}}_{m}\!\!=\!\!\left[\!\!\!\begin{array}{ccccc}\!\mathbf{H}_{2m-1}^{T}\mathbf{B}_{tm}\mathbf{T}_{m}\mathbf{B}_{rm}\mathbf{H}_{2m-1}& \!\!\!\mathbf{H}_{2m-1}^{T}\mathbf{B}_{tm}\mathbf{T}_{m}\mathbf{B}_{rm}\mathbf{H}_{2m} \\  \mathbf{H}_{2m}^{T}\mathbf{B}_{tm}\mathbf{T}_{m}\mathbf{B}_{rm}\mathbf{H}_{2m-1} &\!\!\! \mathbf{H}_{2m}^{T}\mathbf{B}_{tm}\mathbf{T}_{m}\mathbf{B}_{rm}\mathbf{H}_{2m}\end{array}\!\!\!\!\right]\nonumber.\end{align}
{The off-diagonal terms of the matrix $\widehat{\mathbf{H}}_{m}$ are the effective channel between $(2m,2m-1)$ and $(2m-1,2m)$ transmit-receive user-pairs, respectively. Specifically, we denote $\mathbf{H}_{2m,2m-1}$ as effective channel between $(2m-1,2m)$ transmit-receive user-pair} 
\begin{align}\label{H(2m,2m-1)}
\mathbf{H}_{2m,2m-1}=\mathbf{\widetilde{H}}_{2m}^{T} \mathbf{B}_{tm}\mathbf{T}_{m} \mathbf{B}_{rm} \mathbf{\widetilde{H}}_{2m-1}.
\end{align}
We note that $(k,k^{'})$ denotes the user-pair that exchanges information, $(k,k^{'})=(2m-1,2m) \mbox{ or } (2m,2m-1)$ for $m=1,\cdots,K$. With the above design which only cancels the inter user-pair interference and not the self-interference, the~signal received at the user-pair $(k,k^{'})$ can be expressed using \eqref{yk}~as 
\begin{align}\label{yk,yk'}
& \mathbf{y}_{k}= \alpha \mathbf{H}_{k,k} \mathbf{x}_{k} +\alpha \mathbf{H}_{k,k^{'}} \mathbf{x}_{k^{'}} + \mathbf{\widetilde{n}}_{k},\nonumber \\ &\mathbf{y}_{k^{'}}= \alpha \mathbf{H}_{k^{'},k^{'}} \mathbf{x}_{k^{'}} +\alpha \mathbf{H}_{k^{'},k}\mathbf{x}_{k} + \mathbf{\widetilde{n}}_{k^{'}}.
\end{align}
Since each user knows its self-data, it can cancel the self-interference. The resultant signal now is
\begin{align}\label{si_can_eq}
 \tilde{\mathbf{y}}_{k}&=  \alpha \mathbf{H}_{k,k^{'}} \mathbf{x}_{k^{'}} + \mathbf{\widetilde{n}}_{k}= \alpha \sqrt{\frac{p_{k^{'}}}{N_{U}}}\mathbf{H}_{k,k^{'}} \mathbf{D}_{k^{'}} \mathbf{s}_{k^{'}} + \mathbf{\widetilde{n}}_{k},\nonumber \\ 
\tilde{\mathbf{y}}_{k^{'}}&=\alpha \mathbf{H}_{k^{'},k}\mathbf{x}_{k} + \mathbf{\widetilde{n}}_{k^{'}}=\alpha \sqrt{\frac{p_{k}}{N_{U}}}\mathbf{H}_{k^{'},k}\mathbf{D}_{k} \mathbf{s}_{k}+ \mathbf{\widetilde{n}}_{k^{'}}.
\end{align}

\subsection{Design of relay amplification matrix $\mathbf{T}$}
We now design the relay amplification matrix $\mathbf{T}=\text{blkdiag}\left[ \mathbf{T}_{1},\cdots,\mathbf{T}_{K}\right]$, where $  \mathbf{T}_{m} \in \mathbb{C}^{2N_{U} \times 2N_{U}}$ is the  relay amplification matrix for the $m$th user pair. 
We use the algebraic norm-maximizing (ANOMAX) strategy which maximizes the Frobenius norm of the effective channel matrix among the user pairs~\cite{RoemerH09}.  The block-diagonalization combined with ANOMAX transmit strategy  provides the best balance between complexity and performance \cite{icasspZhangRH11}.
The ANOMAX algorithm has the following objective~\cite{RoemerH09}
\begin{subequations}\label{prob}
\begin{alignat}{2}
 &\underset{\mathbf{T}_{m}}{\mbox{ arg max }}
\mathbf{J}_{\beta}(\mathbf{T}_{m})
\label{problem1_1}\\
&\text{ s.t. } {\Vert \mathbf{T}_{m} \Vert_{\text{F}}=1},\label{problem1_2}
\end{alignat}
\end{subequations}
where $\mathbf{J}_{\beta}(\mathbf{T}_{m})\!=\!{\beta^{2} \Vert \mathbf{H}_{2m-1,2m} \Vert_{\text{F}}^{2} \!+\! (1\!-\!\beta)^{2}  \Vert \mathbf{H}_{2m,2m-1} \Vert_{\text{F}}^{2}}$, and the scalar $ \beta \in [0,1] $ is a weighting factor. If we denote $ \mathbf{t}_{m}\triangleq \text{vec}\left\lbrace \mathbf{T}_{m}\right\rbrace  $ and 
\begin{align}\label{L_betam}
\mathbf{L}_{\beta m} & \triangleq \left[ \beta (\mathbf{B}_{rm} \mathbf{\widetilde{H}}_{2m})\otimes(\mathbf{B}_{tm}^{T} \widetilde{\mathbf{H}}_{2m-1})\right.,\nonumber \\
&\qquad\left.(1-\beta)((\mathbf{B}_{rm} \widetilde{\mathbf{H}}_{2m-1})\otimes(\mathbf{B}_{tm}^{T} \widetilde{\mathbf{H}}_{2m})\right], 
\end{align} then 
$\mathbf{J}_{\beta}(\mathbf{T}_{m})=\|\mathbf{L}_{\beta m}^{T}\mathbf{t}_{m}\|$ \cite{RoemerH09}. Using the above transformation, the optimization in \eqref{prob} can be equivalently cast as
\begin{subequations}\label{prob1}
\begin{alignat}{2}
 &\underset{\mathbf{t}_{m}}{\mbox{ max }}
\|\mathbf{L}_{\beta m}^{T}\mathbf{t}_{m}\|
\label{problem1_2}\\
&\text{ s.t. } {\Vert \mathbf{t}_{m} \Vert_{2}=1}.\label{problem1_2}
\end{alignat}
\end{subequations}
%
We note that the above problem is the Rayleigh quotient and its solution is given as the largest singular value of $\mathbf{L}_{\beta m}$, which can be found using its SVD i.e., $\mathbf{L}_{\beta m}=\mathbf{U}_{\beta m}\mathbf{\Sigma}_{\beta m}\mathbf{V}_{\beta m}^{H}$.We therefore choose the optimal vector as $ \mathbf{t}_{m}=\mathbf{u}_{\beta m,1}^{\ast} $, which is the first column of $\mathbf{U}_{\beta m}^{*}$, and consequently the conjugate of the first dominant singular vector of $\mathbf{L}_{\beta m}$. The optimal relay amplification matrix for the $m$th user-pair is consequently given as 
\begin{align}\label{Tm}
\mathbf{T}_{m}=\text{unvec}_{2N_{U}\times 2N_{U}}\left\lbrace \mathbf{u}_{\beta m,1}^{\ast} \right\rbrace.
\end{align}


\subsection{Processing at the users}\label{Proc_at_user}
We now design the baseband precoding and decoding matrices at the users, and the objective is to maximize the multiplexing gain. To achieve this objective, we first whiten the colored noise  at the $k$th terminal {in \eqref{si_can_eq}}, which has covariance matrix $ \mathbf{K}_{z}=\mathbb{E}\left\lbrace \mathbf{\widetilde{n}}_{k} \mathbf{\widetilde{n}}_{k}^{H} \right\rbrace$. We design the whitening filter by computing the eigen value decomposition of $\mathbf{K}_{z}$ which is given as $ \mathbf{K}_{z} =\mathbf{U}_{z} \mathbf{\Sigma}_{z} \mathbf{U}_{z}^{H}$. The whitening filter is 
\begin{align}\label{Kw}
\mathbf{K}_{w}=\mathbf{\Sigma}_{z}^{-\frac{1}{2}} \mathbf{U}_{z}^{H}.
\end{align}
{The whitened end-to-end channel between the $(k,k')$ user-pair, from \eqref{si_can_eq}, can be expressed as  
\begin{align}\label{}
\mathbf{K}_{w}\mathbf{y}_{k}&=  \alpha \sqrt{\frac{p_{k^{'}}}{N_{U}}}\mathbf{K}_{w}\mathbf{H}_{k,k^{'}}  \mathbf{D}_{k^{'}}\mathbf{s}_{k^{'}} + \mathbf{K}_{w}\mathbf{\widetilde{n}}_{k}\nonumber\\
&=\sqrt{\frac{p_{k^{'}}}{N_{U}}} \mathbf{\widehat{H}}_{k,k^{'}} \mathbf{D}_{k^{'}}\mathbf{s}_{k^{'}} + \mathbf{\widehat{n}}_{k}.
\end{align}
}
{
Here $\mathbf{\widehat{H}}_{k,k^{'}}= \alpha \mathbf{K}_{w} \mathbf{H}_{k,k^{'}}$ is the effective channel after whitening, and $\mathbf{\widehat{n}}_{k}= \mathbf{K}_{w}\mathbf{\widetilde{n}}_{k}$ is the whitened noise. We now diagonalize the above channel by performing
SVD of $\mathbf{\widehat{H}}_{k,k^{'}}=\mathbf{\widehat{U}}_{k,k^{'}} \mathbf{\widehat{\Sigma}}_{k,k^{'}} \mathbf{\widehat{V}}_{k,k^{'}}^{H}$,  and then design the baseband precoding and decoding matrices as 
\begin{align}\label{Dk}
\mathbf{D}_{k'}=\mathbf{\widehat{V}}_{k,k^{'}}\mathbf{\overline{\Sigma}}_{k,k^{'}}^{\frac{1}{2}} \mbox{ and } \mathbf{Q}_{k}=\mathbf{\widehat{U}}_{k,k^{'}}^{H}\mathbf{K}_{w},
\end{align}
where $\mathbf{\overline{\Sigma}}_{k,k^{'}}$ is the diagonal power allocation matrix calculated using water-filling algorithm \cite{Paulraj}.
}

\subsection{Amplification factor}
We now compute the amplification factor $\alpha$ which can be written, using \eqref{hatr}, as
\begin{align}\label{alpha}
\!\!\!\!\!\alpha\!=\!\sqrt{\!\frac{P_{R}}{\mbox{Tr}\left\lbrace \widetilde{\mathbf{W}}\left(\sum\limits_{k=1}^{2K}p_{k}\mathbf{H}_{k} \mathbf{D}_{k}\mathbf{D}_{k}^{H}\mathbf{H}_{k}^{H}+ \sigma_{R}^{2}\mathbf{I}_{N_R} \right)\widetilde{\mathbf{W}}^{H}\right\rbrace}}.\!
\end{align}
We see from \eqref{Dk} and \eqref{alpha} that the design of $\mathbf{D}_{k'}$ and $\mathbf{Q}_{k}$, and calculation of $\alpha$ are coupled together.
We, therefore, cannot calculate $\alpha$ using \eqref{alpha} alone. To calculate $\alpha$, and design $\mathbf{D}_{k'}$ and $\mathbf{Q}_{k}$ simultaneously,  we propose an iterative algorithm (see Algorithm~\ref{algo1}).
\begin{algorithm}
\DontPrintSemicolon 
\KwIn{ Set $\alpha^{(0)}=1$, which implies $ \mathbf{W}^{(0)}=\widetilde{\mathbf{W}} $  from \eqref{W}, and maximum iteration as $N_{max}$.}
\For{$n =1$ \textbf{to} $N_{max}$}{
Obtain $ \mathbf{D}_{k^{'}}^{(n)}$ and $ \mathbf{Q}_{k}^{(n)}$ using \eqref{Dk} by substituting the value of   $\mathbf{W}^{(n-1)}$ into \eqref{eq_noise}.
\\
Calculate $ \tilde{\alpha}^{(n)} $ by substituting the value of $\mathbf{D}_{k^{'}}^{n}$ and $\mathbf{W}^{(n-1)}$ in \eqref{alpha}.
\\
Update the value of amplification factor and beamforming matrix as\newline
$\alpha^{(n)}=\alpha^{(n-1)} \tilde{\alpha}^{(n)}$ and 
$ \mathbf{W}^{(n)}=\alpha^{(n)}\widetilde{\mathbf{W}}. $
}
\Return{$\alpha^{(N_{max})}, \mathbf{D}_{k^{'}}^{(N_{max})} , \mathbf{Q}_{k}^{(N_{max})}$.}\;
\caption{Calculation of amplification factor at the relay, and baseband precoding and decoding matrices at the users}\label{algo1}
\end{algorithm}
\vspace{-0.1in}
\section{Simulation results}\label{simulation}
We now numerically investigate the SE of the proposed design, that is calculated using \eqref{SE}. To benchmark the SE of the proposed block-diagonalization-based hybrid processing relay (HPR), we compare its SE with a full RF chain relay (FRR) system. For this study, we set the noise variances as $\sigma_{R}^{2}=\sigma_{k}^{2}=1$, and define SNR $= P_{R}=p_{k}$. We also set the weighting factor $\beta$ to be $0.5$. The simulation results are obtained by averaging over $1000$ statistically independent channel realizations.
\iftoggle{SINGLE_COL}{\begin{figure}[htbp]}{\begin{figure}[htbp]}
    \centering
    \includegraphics[width=\iftoggle{SINGLE_COL}{.45\linewidth}{0.75\linewidth}]{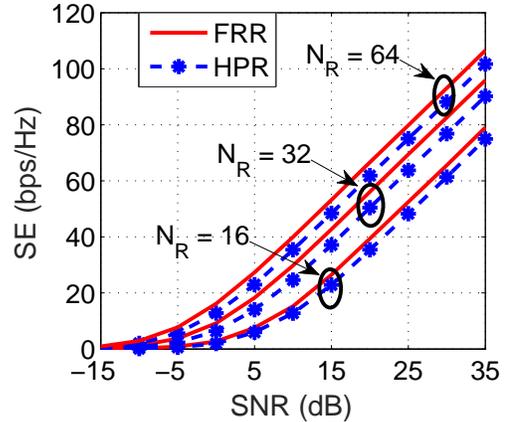}\vspace*{-8pt}
  \caption{\small SE versus SNR for $ K=4$ and $ N_{U}=2. $
   }
    \label{SEvsSNR_diff_NR}
  \end{figure}

We first plot in Fig.~\ref{SEvsSNR_diff_NR} the SE versus SNR of HPR with $K=4$ user pairs, $N_{U}=2$ user antennas and $M_{D}=2$ data streams. We plot these curves for three different values of relay antennas $N_R$. We see that the SE of the both system increases with increase in $N_{R}$, due to increased beamforming gains. We also observe  that both HPR and FRR have similar SE. We next plot in Fig.~\ref{SEvsNR_diff_NU} the SE versus $N_{R}$ for $K=4$ and $\mbox{SNR}=20$~dB for $N_{U}=M_{D}=2,4 \text{ and } 8$ user antennas and data streams. We observe that the SE increases with the increase in $N_{U}$. This is because of increased multiplexing gain due to increased $N_U$, and consequently increased $M_D$.


\iftoggle{SINGLE_COL}{\begin{figure}[htbp]}{\begin{figure}[htbp]}
    \centering
    \includegraphics[width=\iftoggle{SINGLE_COL}{.45\linewidth}{0.75\linewidth}]{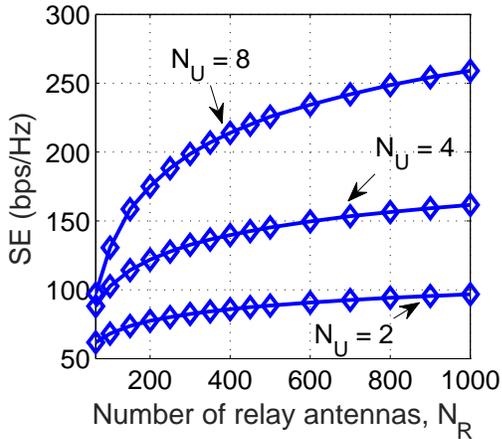}\vspace*{-0pt}
  \caption{\small SE versus number of relay antennas, $N_{R}$ for $K=4$ and $\mbox{SNR}=20$~dB.}
    \label{SEvsNR_diff_NU}
  \end{figure}

\iftoggle{SINGLE_COL}{\begin{figure}[htbp]}{\begin{figure}[htbp]}
    \centering
    \includegraphics[width=\iftoggle{SINGLE_COL}{.45\linewidth}{0.75\linewidth}]{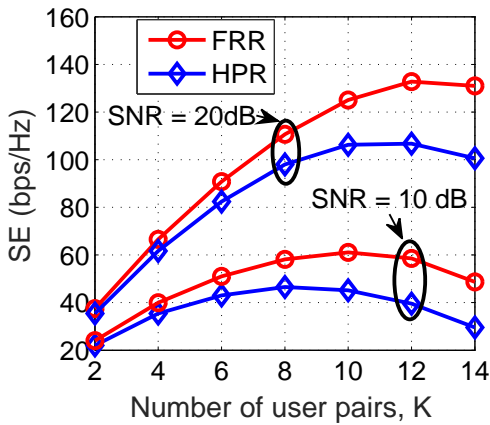}\vspace*{-0pt}
  \caption{\small SE versus number of user pairs, $K$ for $ N_{R}=64 $ and $ N_{U}=2$. }
    \label{SEvsK_diff_SNR}
  \end{figure}

We now plot in Fig.~\ref{SEvsK_diff_SNR} the SE versus $K$ with $N_{R}=64$ and $N_{U}=M_{D}=2$, for two different SNR values. We see that the difference in the SE of HPR and FRR systems increases with increase in the number of user-pairs $K$. This is because a FRR system has better ability to cancel increased inter-user interference (due to increased number of user-pairs), and can crucially provide higher beamforming gains. We notice that the SE of both HPR and HRR systems initially increases with increase in $K$, and then reduces.  This is because at low SNR values of $10$ and $20$~dB (considered in the plots), the block-diagonalization design, while cancelling the inter user-pair interference, also boosts the noise. This noise boost is conspicuously higher for large $K$ values, which  degrades the SE of both HPR and FRR.

\iftoggle{SINGLE_COL}{\begin{figure}[htbp]}{\begin{figure}[htbp]}
    \centering
    \includegraphics[width=\iftoggle{SINGLE_COL}{.45\linewidth}{0.75\linewidth}]{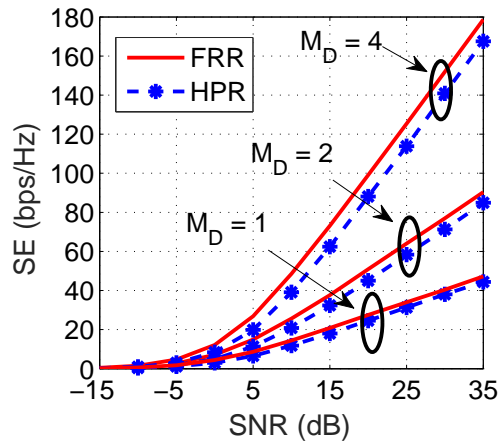}\vspace*{-0pt}
  \caption{\small SE versus SNR for $N_{R}=64, K=4  $ and $ N_{U}=4 $.}
    \label{SEvsSNR_diff_MD}
  \end{figure}

We finally plot in Fig.~\ref{SEvsSNR_diff_MD} the SE versus SNR with $K=4$ and $N_{U}=4$ for different $M_{D}$ values, which denotes the number of data streams per user. We see that, due to increase in multiplexing gain  with increase in $M_D$, the SE for both FRR and HPR increases, which is not surprising. More importantly, with increase in $M_D$ values, the gap between the SE of two systems increases slightly. This is again because an FRR system  i) has better inter-user interference cancelling ability; and ii)  provides better beamforming gain. 



\section{Conclusion}\label{conclusion}
We constructed a hybrid design for multi-pair two-way HD massive MIMO relaying with MIMO users. The proposed design uses block-diagonalization approach in the digital baseband domain, and equal-gain-combing approach in the analog RF domain. The approach decouples a multiple user-pair two-way hybrid relaying system into interference-free multiple single user-pair two-way hybrid relaying systems. We numerically demonstrated that the spectral efficiency of the proposed design is close to that of a full-RF chain relaying system, and that too with lesser implementation complexity.
\bibliographystyle{IEEEtran}
\bibliography{IEEEabrv,hybrid}

\end{document}